# LLMs: A Game-Changer for Software Engineers?

**Md. Asraful Haque**

**(ORCID: 0000-0002-0518-2785)**

Computational Unit, Z.H. College of Engineering & Technology
Aligarh Muslim University, Aligarh-202002, India
Email: md_asraf@zhcet.ac.in

## Abstract

Large Language Models (LLMs) like GPT-3 and GPT-4 have emerged as groundbreaking innovations with capabilities that extend far beyond traditional AI applications. These sophisticated models, trained on massive datasets, can generate human-like text, respond to complex queries, and even write and interpret code. Their potential to revolutionize software development has captivated the software engineering (SE) community, sparking debates about their transformative impact. Through a critical analysis of technical strengths, limitations, real-world case studies, and future research directions, this paper argues that LLMs are not just reshaping how software is developed but are redefining the role of developers. While challenges persist, LLMs offer unprecedented opportunities for innovation and collaboration. Early adoption of LLMs in software engineering is crucial to stay competitive in this rapidly evolving landscape. This paper serves as a guide, helping developers, organizations, and researchers understand how to harness the power of LLMs to streamline workflows and acquire the necessary skills.

## 1. Introduction

Software engineering (SE) processes refer to the structured set of activities involved in the development of software systems, including requirements analysis, design, coding, testing, deployment, and maintenance. These processes ensure that software is built systematically and meets user needs while maintaining quality and reliability [1]. Software development follows various models such as the Waterfall, Agile, or DevOps, each outlining different approaches to these phases. Software engineering can be costly and time-consuming for several factors related to the complexity, labor intensity, and long-term maintenance requirements. Fig.1 illustrates a typical breakdown of the effort or cost allocation throughout the various phases of software development life cycle (SDLC) [2][3]. The primary objective of software engineering is to develop high-quality software at a minimal cost. The software industry faces numerous challenges in developing reliable software, particularly as systems become increasingly complex [4]. The demand for faster development cycles, high-quality code, and the ability to handle large-scale systems has driven the adoption of new tools and technologies. Among these, Large Language Models (LLMs) have emerged as a powerful force, automating and optimizing various aspects of the software engineering process [5]. Large language models are state-of-the-art NLP tools that have been trained on massive amounts of data, allowing them to generate human-like responses and understand complex language patterns. They have gained immense popularity in recent years because it makes a lot of things easier and quicker. They have the potential to revolutionize various industries and transform the way we interact with technology. They have demonstrated impressive capabilities that are directly applicable to software engineering [6][7]. Some of the key functions include code generation, debugging, testing etc. The integration of Large Language Models (LLMs) into software engineering (SE) is transforming traditional practices in multiple ways. From altering how developers write, review, and maintain code to revolutionizing collaboration within teams, LLMs are reshaping the landscape of SE [8-11]. The impact of LLMs on software engineering tools and platforms is evident in the growing trend of LLM-powered IDEs. These environments now offer intelligent code suggestions, natural language queries, and automated refactoring, making development more intuitive. While there's much excitement about LLMs in software engineering, significant concerns remain regarding their practical use and ethical implications [12]. LLMs lack true comprehension of the logic behind code, making them prone to generating incorrect or insecure outputs. Additionally, the adoption of these models also brings challenges related to ethics, job roles, and the need for careful human oversight. Thus the question remains: Are these capabilities sufficient to significantly transform the software engineering industry?

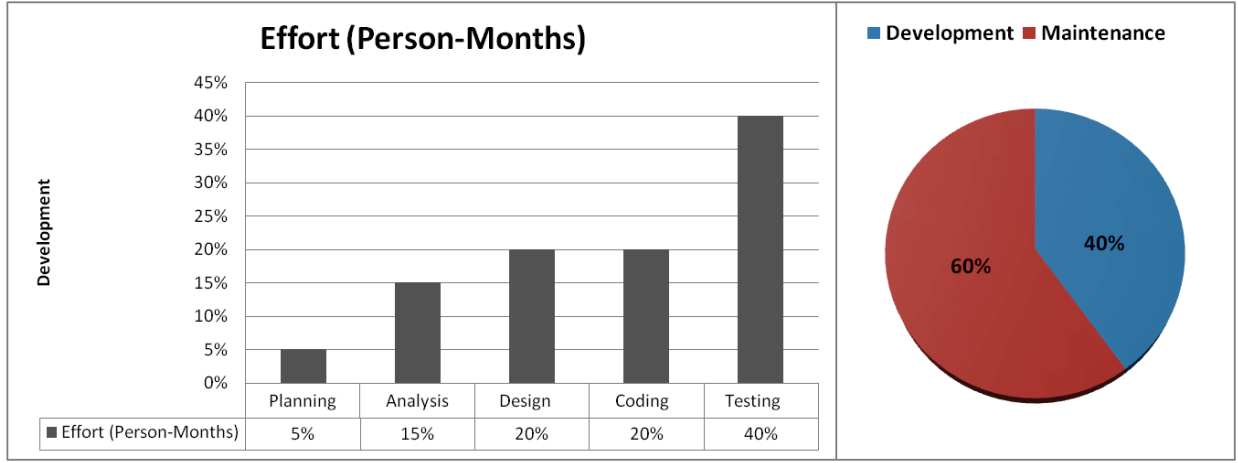

Fig 1. Typical effort distribution at different phases of SDLC

In this paper, we aim to explore the transformative potential of Large Language Models in software engineering, assessing whether they represent an overhyped trend or a disruptive innovation capable of reshaping the field. We will delve into the technical strengths and limitations of LLMs, examine real-world case studies, and discuss the ethical considerations that come with the adoption of AI-driven development tools. Through this comprehensive analysis, we seek to provide a balanced perspective on the role of LLMs in modern software engineering practices.

## 2. Understanding Large Language Models

Large language models (LLMs) are built on the transformative power of the transformer architecture, a model introduced by Vaswani et al. in 2017 [13] that has since become the foundation of many advanced LLMs. The transformer architecture, unlike its predecessors like recurrent neural networks (RNNs) and long short-term memory (LSTM) networks, excels at handling long-range dependencies in data through its self-attention mechanism. This mechanism enables the model to understand and weigh relationships between all tokens in a sequence simultaneously, rather than processing them in order. This ability to capture both local and global context makes transformers highly effective for tasks that require understanding the structure and flow of text or code. In the context of software engineering, this allows LLMs to not only generate code based on natural language prompts but also to understand the intricate relationships between different parts of a codebase, which is crucial for complex tasks like debugging, code completion, and refactoring. The self-attention mechanism is a key innovation that empowers LLMs to efficiently determine the importance of different parts of input data, whether in a sentence or in a block of code. This helps LLMs better understand the context of programming languages, allowing them to predict the next steps in coding processes or provide useful suggestions during the development cycle. Another vital aspect of transformer models is positional encoding, which helps maintain the order of input data — a necessary feature when processing sequences like code, where the position of elements is critical to functionality. The combination of self-attention and positional encoding allows LLMs to process code sequences with an understanding of both immediate context and overall structure, thus improving their performance in code generation and related tasks.

The development of LLMs involves mainly three stages: pre-training, fine-tuning and reinforcement learning with Human Feedback [14-16]. In the pre-training phase, LLMs are exposed to vast amounts of textual and coded data, learning general language patterns, coding structures, and syntax from diverse sources such as books, websites, and open-source code repositories. The scope of this pre-training enables LLMs to acquire a broad understanding of multiple programming languages and frameworks, making them versatile in handling different software engineering tasks. Once pre-training is complete, the model undergoes fine-tuning on specific datasets tailored to the target application, refining its ability to perform tasks in specialized areas such as web development, cybersecurity, or enterprise software solutions. This fine-tuning process sharpens the model's ability to generate relevant, high-quality outputs in response to domain-specific inputs. At the end, reinforcement learning is used to further enhance the model's performance by interacting with an environment and receiving human feedback. The feedback, in the form of ratings, rankings, or corrections, is used as a reward signal to guide the model's learning. This is an iterative process and continues until the model meets the desired standards, at which point it can be used in real-world applications. A typical training process of OpenAI's ChatGPT has been shown in Fig.2 [17]. By leveraging the advantages of pre-training, fine-tuning and RLHF, LLMs become proficient in understanding not only the general syntax and structure of code but also in adapting to specialized coding practices and conventions. This allows LLMs to assist with software engineering tasks such as code generation, debugging, and even testing, making them valuable tools for developers working across a variety of programming languages and problem domains. Despite these strengths, however, LLMs still face challenges, particularly when dealing with complex logic or novel problems outside of their training data. Nevertheless, their advanced architecture and training methodologies have positioned them as powerful, versatile tools in the field of software engineering.

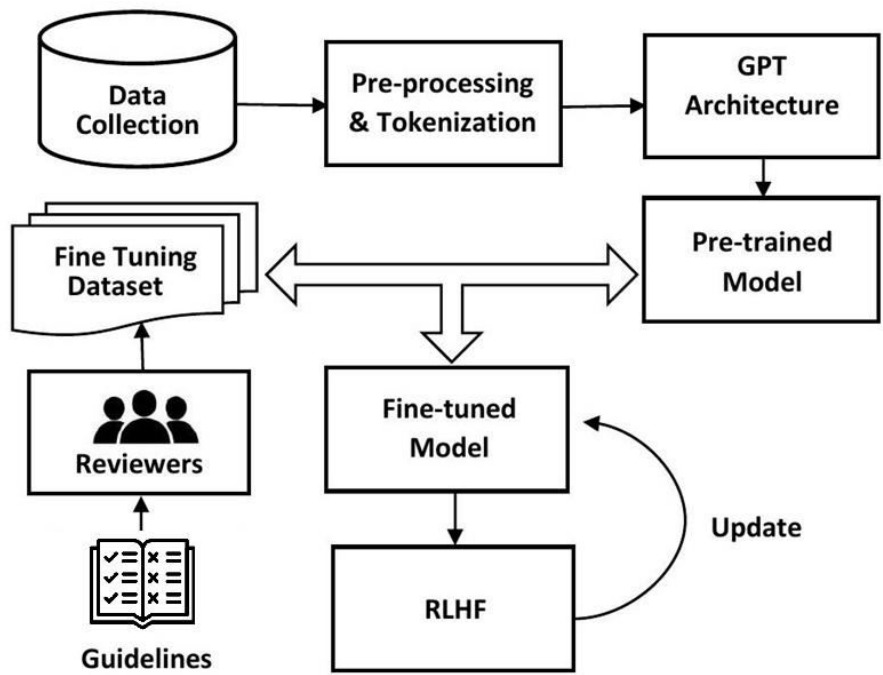

Fig 2. ChatGPT Training Process

A significant number of LLMs are already in use. Table 1 provides a brief overview of some well-known models [17-22]. The future holds promise for even more powerful and advanced LLMs.

Table 1. A brief history of some prominent LLMs

| LLMs | Release Date | Developer | Model Size | |
|------|-------------|-----------|-----------|---|
| | | | **Number of Parameters** | **Dimension (L×H)\*** |
| BERT | October-2018 | Google | 110 billion (Base Model) | L=12, H=12 (Base Model)\*\* |
| GPT-2 | February-2019 | OpenAI | 1.5 billion | L=12, H=12 (Small Version)\*\* |
| XLNet | June-2019 | Google & CMU | 110 million (Base Model) | L=12, H=12 (Base Model)\*\* |
| T5 | October-2019 | Google | 11 billion | L=12, H=12 (Small Version)\*\* |
| GPT-3 | June-2020 | OpenAI | 175 billion | L= 96, H= 96 |
| Codex | August-2021 | OpenAI | 12 billion | L= 24, H= 32 |
| PaLM | April-2022 | Google | 540 billion | L= 118, H= 128 |
| GALACTICA | November-2022 | Meta AI | 120-billion | L= 80, H= 96 |
| LLaMA | February-2023 | Meta AI | 65 billion | L= 80, H= 64 |
| GPT-4 | March-2023 | OpenAI | 1.76 trillion | Details undisclosed |
| Gemini 1.5 | May-2024 | Google DeepMind | Details undisclosed | Details undisclosed |

\*L=Number of layers, H=Number of attention heads.

\*\*These models have different variants.

## 3.  Technical Strengths and Benefits of LLMs in SE

Large language models (LLMs) have brought transformative potential to software engineering, providing a suite of technical strengths and benefits that can drastically enhance productivity, code quality, and innovation (Fig.3). From improving code generation to automating complex documentation tasks, LLMs are reshaping how developers approach various phases of the software development lifecycle. Below is a detailed exploration of the technical strengths and benefits of LLMs in software engineering (SE).

### 3.1. Code Generation

One of the most prominent uses is code generation, where models like GitHub Copilot, powered by OpenAI's Codex, allow developers to describe the functionality they need in natural language, and the LLM generates relevant code snippets. This not only speeds up the coding process but also minimizes repetitive tasks, enabling developers to focus on more complex aspects of software design and architecture [23][24]. The benefit here is a marked improvement in productivity, as LLMs assist in automating routine coding activities like writing boilerplate code, implementing standard algorithms, or creating simple data structures. Furthermore, this application is versatile across different programming languages, offering cross-language flexibility that is particularly useful in polyglot development environments where multiple languages are used. The ability to generate code across Python, JavaScript, Java, C++, and other languages adds immense value, reducing the need for developers to switch contexts or master multiple languages to complete tasks efficiently [25-28].

### 3.2. Code Review, Debugging and Testing

LLMs have the ability to automate code reviews and assist with bug detection [29]. These models can facilitate the knowledge required for high-quality code reviews. Even junior developers, with the assistance of LLM-powered tools, can contribute effectively to the code review process by leveraging the model's knowledge of industry standards and best practices. Traditionally, debugging requires manual effort from developers, who inspect the code for errors, and once the error is located developers can then implement a fix to correct the error [30]. LLMs can analyze logs, error messages, and code execution paths to suggest potential causes of bugs [31-34]. This helps developers quickly pinpoint the source of an issue, reducing the time spent on manual debugging. LLMs can also suggest potential fixes based on the patterns they have learned from analyzing similar bugs in the past. This capability is particularly useful in large, complex systems where tracking down bugs can be a challenging and time-consuming task. For example, LLMs can identify common mistakes such as unhandled exceptions, resource leaks, or improper variable initializations. They can also flag potential security issues like injection vulnerabilities, insecure data handling, or incorrect encryption implementations. By catching these issues early, LLMs help developers produce more secure and robust code. Additionally, LLMs have been applied to automated testing, where they generate unit tests, identify edge cases, and suggest test cases based on the functionality described in code [35][36]. This application can significantly speed up the testing phase of software development, ensuring that code is rigorously tested without developers having to manually write every possible test case. LLMs' capacity to generate exhaustive test suites helps in reducing the likelihood of bugs making it to production, thus increasing the overall robustness and reliability of software systems.

### 3.3. Language and Framework Agnostic

LLMs have the remarkable ability to work across a wide variety of programming languages and frameworks, making them versatile tools in multi-language environments [37]. Because LLMs are trained on diverse datasets that include code from many different programming languages, they can switch between languages and frameworks with ease. This is especially useful for developers who work in environments that require knowledge of multiple languages, such as Python for backend services, JavaScript for frontend development, and SQL for database management. By supporting a broad range of languages and frameworks, LLMs eliminate the need for developers to switch between different coding assistants or learn new tools for each technology they use [38-40]. This contributes to a more seamless development experience and enhances overall efficiency.

### 3.4. Refactoring and Optimization

As software systems grow, they often accumulate technical debt, requiring refactoring and optimization to ensure long-term performance and maintainability. LLMs can assist with these processes by suggesting refactoring opportunities, such as code that can be simplified, duplicated code that can be consolidated, or outdated structures that need updating [41]. This is particularly valuable in large codebases where manually identifying areas for refactoring would be time-consuming and prone to oversight. LLMs can also help optimize code by suggesting more efficient algorithms or design patterns based on established best practices. For instance, if a developer writes a brute-force solution for a problem, the LLM might suggest a more optimal approach using dynamic programming or divide-and-conquer algorithms. Additionally, LLMs can provide performance insights, such as identifying inefficient loops, excessive memory usage, or potential bottlenecks in code execution, which helps ensure that the software remains scalable and performant as it evolves [42].

### 3.5. Automated Documentation

Generating accurate, up-to-date documentation has long been a challenge for software engineers, as it is often seen as tedious work that lags behind code changes [43]. Developers often deprioritize this due to tight deadlines or a focus on feature development. However, documentation is crucial for ensuring that code is maintainable, understandable, and transferable across teams and developers. LLMs can analyze code and automatically generate documentation for codebases, including explaining the purpose and functionality of specific functions, classes, and modules [44]. This ensures that documentation stays current and can be updated as code evolves, providing developers with easily understandable, well-structured explanations of how various parts of the system work. This capability not only improves team collaboration and knowledge sharing but also supports on-boarding processes by helping new developers quickly understand legacy codebases or complex systems. In agile environments, where requirements and implementations frequently change, the ability of LLMs to update documentation dynamically as the code evolves is an invaluable benefit.

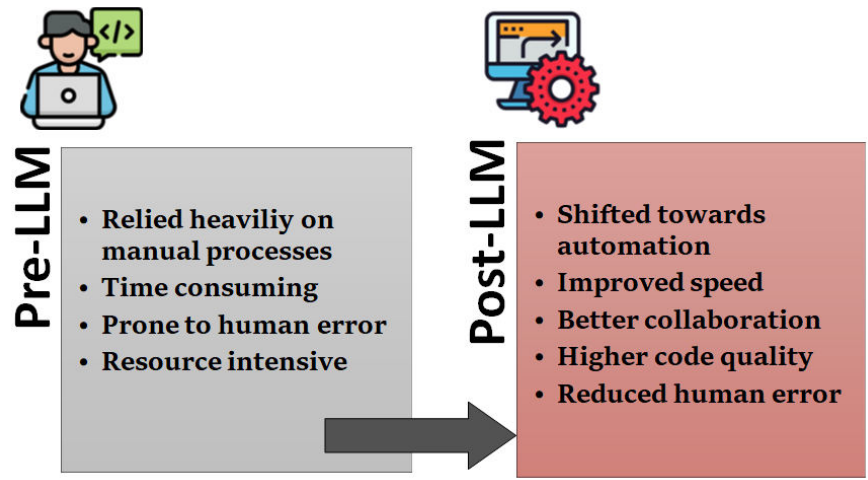

Fig 3. Revolution in SE Practices

## 4. Challenges

While LLMs offer exciting possibilities in the SE domain, several technical limitations and a range of ethical challenges must be addressed:

### 4.1. Technical Limitations

- **Lack of True Understanding**: LLMs do not "understand" code in the same way humans do [8][44]. While LLMs are powerful at predicting sequences based on statistical patterns learned from vast datasets, they lack a deep understanding of the underlying logic and intent behind a given piece of code. In software engineering, this is particularly problematic because coding often requires not just syntactically correct solutions, but solutions that align with specific business logic, system architecture, and performance requirements [45]. For instance, an LLM may generate syntactically correct code for a sorting algorithm but fail to account for efficiency constraints such as time complexity or memory usage, especially when these concerns are implicit in the task description. The inability of LLMs to grasp these nuances means that while they are useful for generating code snippets or suggesting fixes, developers must rigorously review and adapt their outputs to ensure they meet the functional and non-functional requirements of the system.

- **Context Sensitivity**: Although LLMs are good at handling short, localized contexts, they often struggle with maintaining long-term context over extended portions of a codebase [46]. Software systems are often composed of multiple files, modules, and libraries that interact with one another in complex ways. Maintaining context across such a large-scale system, where changes in one part of the codebase can have ripple effects across the system, is a challenge for LLMs. For example, an LLM may generate code that works well within a single function but fails to account for broader architectural considerations, such as how this function interacts with others in different modules or libraries. In large enterprise-scale applications, this limitation becomes even more pronounced, as developers need to track dependencies across various subsystems, which LLMs may not handle effectively. The model's understanding tends to deteriorate when it needs to work with codebases that span across multiple files or projects, resulting in incomplete or incorrect suggestions.

- **Inability to Handle Novel or Rare Problems:** LLMs rely heavily on patterns learned from their training data, which means they perform best when tasked with solving common or well-documented problems. However, when faced with novel or rare problems that deviate from established patterns, LLMs often struggle to produce correct or meaningful output. In software engineering, developers frequently encounter unique challenges that require creative problem-solving and a deep understanding of both the problem domain and system architecture. LLMs, constrained by the limitations of their training data, may not have seen enough similar examples to provide an adequate solution. This is especially true for cutting-edge technologies or innovative software designs that have not been widely adopted and thus are not well-represented in public datasets. Furthermore, rare edge cases, which are often the most critical and challenging parts of software development, tend to be poorly handled by LLMs due to the lack of exposure to similar situations during training.

- **Computational Costs**: Large Language Models (LLMs) are computationally intensive, requiring significant hardware resources for training and inference [47]. Training LLMs requires immense computational resources, including large clusters of GPUs or TPUs, leading to high financial costs. Inference, or using the trained model for tasks, can also be computationally expensive, especially for larger models. These high computational costs can be a barrier to entry for organizations, limiting their ability to adopt and utilize LLMs in their software engineering workflows. Another consideration is the energy consumption associated with running LLMs. The large-scale deployment of LLMs in software engineering environments contributes to increased energy usage, which has both economic and environmental implications [48].

- **Transparency and accountability:** LLMs are often seen as black-box models, meaning that their decision-making processes are not easily interpretable by users [49]. When an LLM generates code, it is not always clear how or why it arrived at a particular solution [50]. This lack of transparency becomes problematic in scenarios where LLMs make critical decisions, such as in safety-critical systems or in applications that have legal and regulatory implications. If a software failure occurs due to an LLM's suggestion, it is difficult to assign responsibility — does the fault lie with the developer, the AI, or the organization that provided the AI? This lack of clear accountability creates challenges in governance and compliance, particularly in regulated industries like finance, healthcare, and transportation. Therefore, ensuring that LLMs are explainable and that there are mechanisms in place to track and audit AI-generated outputs is essential for fostering trust and ensuring that ethical guidelines are followed.

- **Security Risks**: If LLMs are trained on large public datasets, including code repositories, they may inadvertently learn insecure or vulnerable coding practices [51]. For instance, if an LLM is trained on a repository where code contains hard-coded credentials, weak encryption methods, or unpatched vulnerabilities, the model may unknowingly generate code that replicates these flaws. This becomes particularly dangerous in security-critical applications like financial software, healthcare systems, or government infrastructure. The potential for LLMs to suggest insecure code increases the burden on developers to scrutinize the model's output closely, ensuring that it adheres to industry best practices and security standards. Therefore, while LLMs can be helpful for automating routine tasks, they should not be used blindly, especially in areas where security is paramount. Continuous oversight and refinement of the training data, as well as integration with secure coding practices, are essential to mitigate these risks.

## 4.2. Ethical Considerations

- **Copyright and Intellectual Property:** LLMs are trained on publicly available data, but this data may include proprietary or copyrighted code that the model can later reproduce in different contexts [52]. When LLMs generate code that closely resembles or directly replicates code from its training data, it raises serious questions about ownership and accountability. Developers using LLM-generated code may inadvertently violate copyright laws if the generated code mirrors protected material without proper attribution. This could lead to legal disputes and undermine the trust in LLMs as reliable tools in professional software development environments. To address these concerns, companies providing LLM services must implement safeguards that either filter copyrighted material during the training process or ensure that LLM-generated content is appropriately flagged for potential legal issues.

- **Biases in Training Data**: One of the most pressing ethical concerns surrounding LLMs is the issue of bias in training data [53]. LLMs are trained on vast datasets that include both natural language text and code from public repositories, such as GitHub, Stack Overflow, and various forums. However, these sources can contain biased, outdated, or even harmful practices. For example, if the training data includes discriminatory language or biased coding patterns (such as gender or race-based assumptions in user data processing), the LLM may learn and perpetuate these biases in its outputs. In the context of software engineering, biased code generation can lead to inequitable software solutions, unfair user experiences, or even legal and reputational risks for companies. Moreover, models trained on real-world codebases may inherit the biases of past software engineering decisions, such as assumptions about users' technical abilities or geographical location, resulting in software that does not serve all demographics equally. Addressing these biases requires careful dataset curation, as well as developing methods for identifying and mitigating bias in LLM outputs.

- **Impact on the Workforce:** The impact on the workforce is another ethical issue associated with the rise of LLMs in software engineering. By automating tasks like code generation, testing, and debugging, LLMs have the potential to reduce the demand for certain types of coding jobs, particularly entry-level or junior software development roles [17][53]. This could lead to job displacement for new developers or those in low-skilled positions, creating economic inequality within the industry. Additionally, reliance on LLMs may result in a deskilling of the software engineering workforce. If developers become too dependent on AI-generated code and suggestions, they may lose the ability to write complex code or troubleshoot issues independently. This could diminish the overall expertise within the field over time, affecting the quality of software and innovation. To counteract these risks, educational systems and organizations need to evolve, focusing on upskilling developers to work alongside LLMs rather than being replaced by them. Training programs should emphasize higher-order skills like software architecture, algorithm design, and critical thinking, which cannot be easily replicated by AI.

## 5. Recent Trends and Case Studies

Software industries worldwide are leveraging AI tools to streamline processes, increase efficiency, and foster creativity in problem-solving. The AI Index 2024 Annual Report [54] highlights that software developers are among the professionals most likely to incorporate AI in their work. As AI's role within the economy grows, understanding how developers use and view AI is becoming essential. Stack Overflow, the Q&A platform for programmers, runs an annual survey targeting developers. For the first time in 2023, this survey gathered insights from over 90,000 developers — featured questions on usage of AI tools. It explored how developers employ these tools, which ones they prefer, and their overall perceptions of them. Table 2 shows the developers' preferences for using AI tools in software engineering tasks. Fig.4 is the graphical representation of Table 2.

Table 2. Developers' preferences for using AI tools

| Development Tasks | Currently using | Interested in using | Not interested |
|---|---|---|---|
| Planning | 13.52% | 38.54% | 29.77% |
| Coding | 82.55% | 23.72% | 4.48% |
| Code Reviews | 10.09% | 49.51% | 22.95% |
| Debugging | 48.89% | 40.66% | 6.37% |
| Testing | 23.87% | 55.17% | 11.44% |
| Documentation | 34.37% | 50.24% | 8.07% |
| Maintenance | 4.74% | 45.44% | 28.33% |
| Learning Codebase | 30.10% | 48.97% | 13.09% |
| Collaboration | 3.65% | 29.98% | 41.38% |

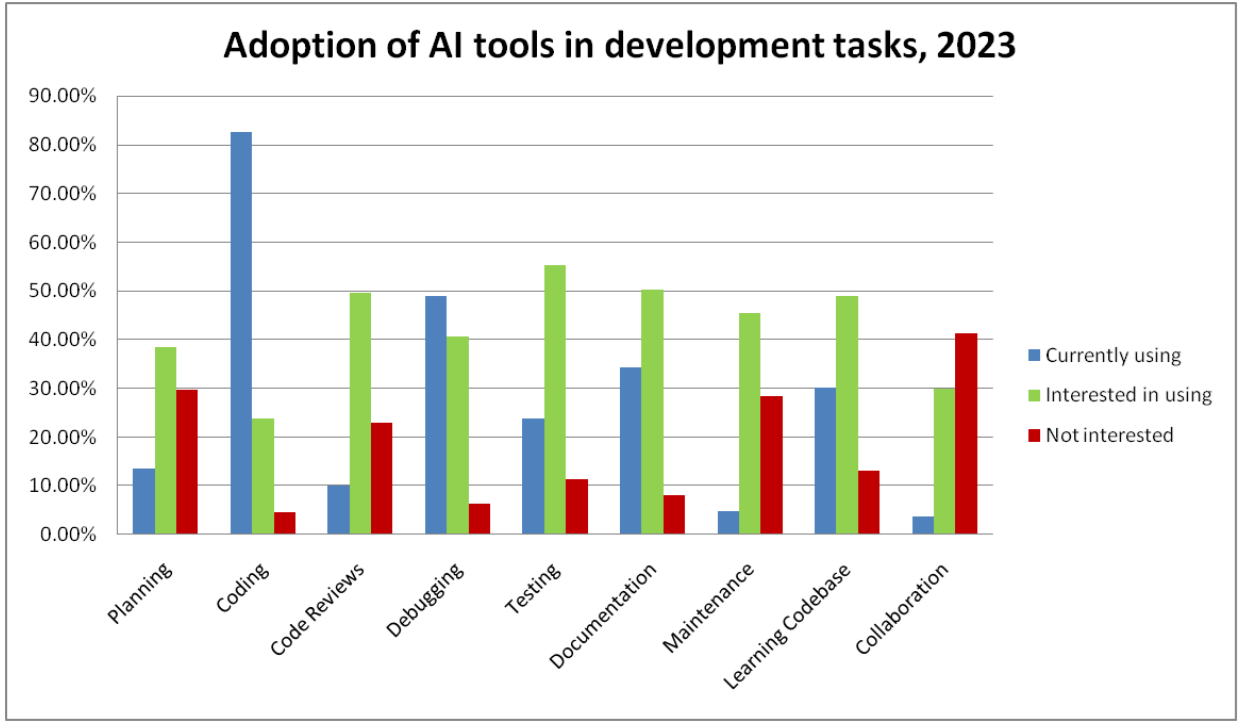

Fig 4. Developers' preferences for using AI-tools in development tasks, 2023

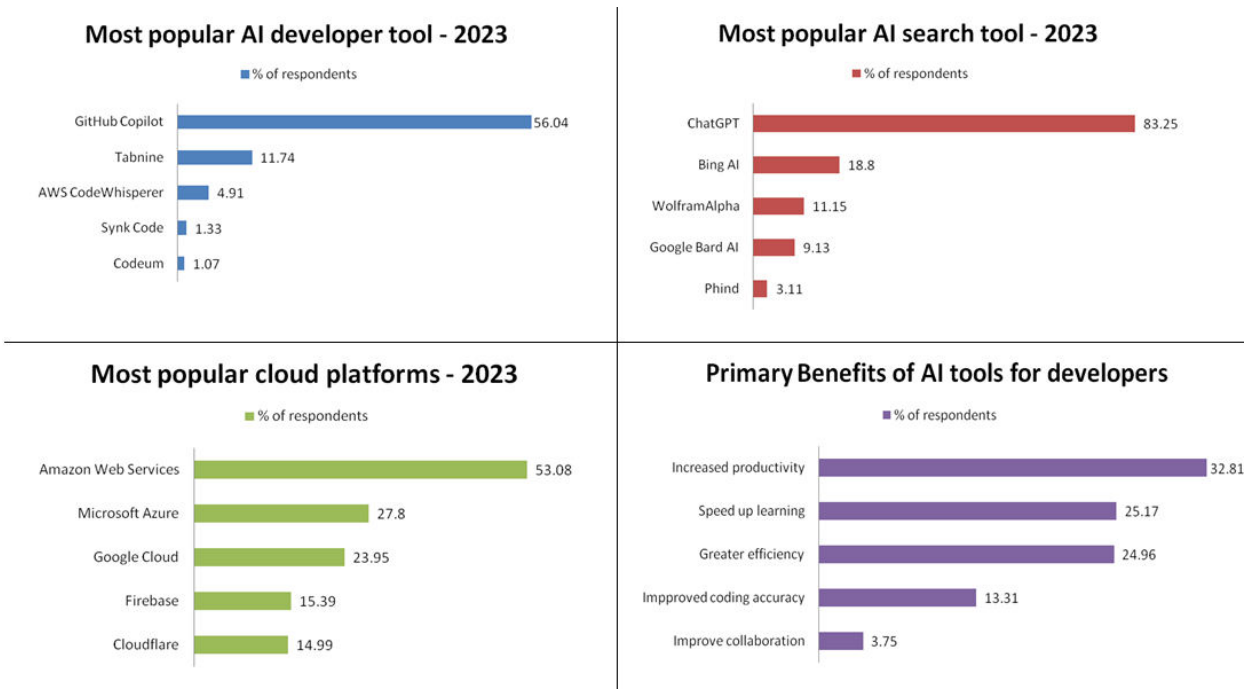

Fig 5. Popularity of AI-tools among professional developers, 2023

The survey was taken in May 2023, thus it may not reflect the availability of more recent AI technologies such as Gemini and Claude 3. The other findings of that survey were as follows (Fig.5):

- Most popular AI developer tool among professional developers, 2023 is GitHub Copilot.
- Most popular AI search tool among professional developers, 2023 is ChatGPT.
- Most popular cloud platform among professional developers, 2023 is Amazon Web Services.
- Developers cited higher productivity (32.8%), quicker learning (25.2%), and increased efficiency (25.0%) as the top benefits of AI tools in their work.

GitHub also conducted a survey [55] from February 26 to March 18, 2024, among 2,000 non-student, corporate respondents in the United States, Brazil, India, and Germany who are not managers and work for organizations with 1,000 or more employees. According to the survey, developers are increasingly integrating AI tools, with the majority of respondents reporting that AI improves their productivity and coding skills. Fig.6 represents the respondents view on the benefits of AI tools. It highlights that popularity and use of AI tools varies by region. Fig.7 displays the current usage of AI coding tools against the corporate endorsement for AI-driven coding. The survey respondents reported that AI tools boost productivity, freeing them up to focus on strategic tasks like system design and client collaboration. To fully leverage AI, organizations should integrate it into every phase of development. AI isn't a job replacement but an enhancer of human creativity.

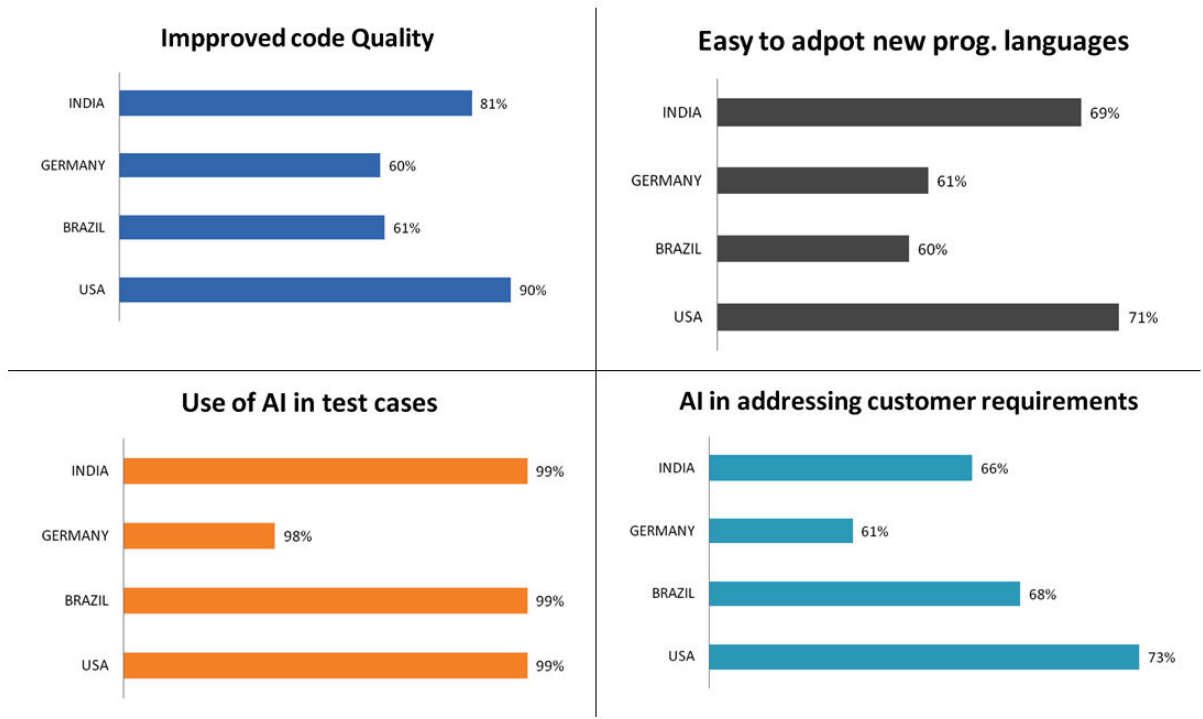

Fig 6. Respondents view on the benefits of AI tools

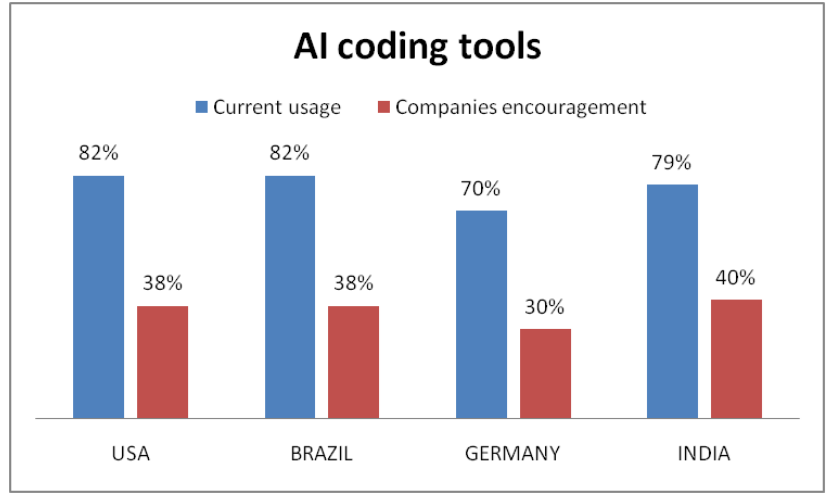

Fig 7. Usage of AI tools vs. Companies encouragement

Initially, AI tools were primarily used for specific tasks like code completion and bug detection. However, the advent of Large Language Models (LLMs) has ushered in a new era of AI-powered development. This shift promises not only greater efficiency but also new possibilities in adaptive, responsive coding environments. In exploring the impact of large language models (LLMs) on software engineering, it is essential to examine real-world case studies where LLMs have been applied in different software engineering (SE) environments. By analyzing diverse cases, we can understand how LLMs can be a game-changer in some situations, while potentially overhyped or insufficient in others.

### 5.1. Microsoft's GitHub Copilot

Microsoft's GitHub Copilot, powered by OpenAI's Codex, is an LLM-based tool integrated into GitHub, one of the largest code hosting platforms used by enterprises globally. It is designed to assist developers by offering code suggestions, autocompletion, and error detection [56-58]. It seamlessly integrates into popular IDEs like Visual Studio Code. Microsoft and GitHub have used Copilot internally, and in collaboration with external developers, to accelerate coding, especially for repetitive tasks and boilerplate code.

- **Key Impacts:**

(i) Copilot saves time on basic coding jobs by offering autocomplete recommendations, allowing developers to focus on high-level reasoning rather than syntax.

(ii) Copilot acts as a mentor, accelerating the learning curve for junior developers by suggesting best practices in various programming languages.

(iii) Copilot promotes code quality by suggesting well-tested, reusable code components, which reduces the need for extensive code review.

- **Challenges:**

(i) Concerns have been raised about the development of unsafe code, since Copilot may recommend code with vulnerabilities if it learns from incorrect sources [59].

(ii) Since the model was trained on publicly available code, companies must consider the potential legal and ethical implications of using its generated code in proprietary systems.

### 5.2. Salesforce's CodeGen

Salesforce developed CodeGen, an advanced language model designed for code generation and understanding. It's part of Salesforce's AI Research initiative and is open-source, making it accessible to a wide range of developers. By leveraging the power of AI, CodeGen aims to accelerate the software development process and enhance developer productivity [60][61].

- **Key Impacts:**

(i) CodeGen allows Salesforce engineers to rapidly generate customization scripts based on user specifications, cutting down hours of manual coding.
(ii) The tool helps consultants quickly adapt to client requirements, allowing for more scalable and customized solutions.
(iii) Engineers can use CodeGen to generate prototypes quickly, leading to faster iteration cycles during the proposal stage of software development.

- **Challenges:**

(i) Generated code often requires significant refactoring, especially for complex client requirements.
(ii) The model occasionally generates code that works well in the short term but may lead to maintenance problems in the long run, compromising software quality.

### 5.3. Meta's TestGen-LLM

Meta's TestGen-LLM is an advanced AI model designed to generate unit tests for software, enhancing code quality and speeding up the testing phase in development. Leveraging its understanding of code patterns, TestGen-LLM suggests relevant test cases, automates repetitive testing tasks, and provides insights into potential vulnerabilities, making it a powerful aid for developers aiming to improve software reliability [62][63].

- **Key Impacts:**

(i) TestGen-LLM helps to improve the overall coverage of the codebase by generating new test cases.
(ii) The generated tests help to uncover and fix potential issues in the code.
(iii) By automating testing tasks, TestGen-LLM saves the time required to release new features.

- **Challenges:**

(i) For highly complex codebases with intricate logic, TestGen-LLM may struggle to generate comprehensive and effective test cases.
(ii) The quality of the generated test cases is heavily reliant on the quality of the existing codebase. If the code is poorly structured or has many dependencies, TestGen-LLM may generate less effective tests.

### 5.4. ChatGPT for Software development

ChatGPT has rapidly become a versatile tool in the software development field, offering assistance across the entire development lifecycle [64]. From ideation and code generation to debugging and documentation, ChatGPT acts as a virtual assistant that accelerates the pace of development. Trained on a wide range of programming languages, software development principles, and technical resources, it integrates seamlessly into workflows, helping individual developers and teams increase efficiency and solve technical issues in real time. ChatGPT's application ranges from rapid prototyping to complex architectural advice, making it a valuable resource in both agile and traditional development environments [65-68].

- **Key Impacts**

(i)     ChatGPT streamlines code generation and refactoring by providing reusable snippets, autocompletion, and efficient solutions for common tasks, leading to faster development cycles and reduced manual coding.

(ii)    With real-time analysis and interpretation of error messages and stack traces, ChatGPT helps developers troubleshoot issues quickly, offering alternative solutions and relevant documentation to address bugs and improve code quality.

(iii)   ChatGPT can generate clear and concise documentation for code, improving code maintainability and collaboration.

(iv)    Developers can use ChatGPT to learn new programming languages and frameworks by asking questions and experimenting with code. It is very useful for new developers, clearing their doubts and helping them understand codebases and workflows more efficiently

- **Challenges**

(i)     ChatGPT may lack the full context of a project, leading to suggestions that might not fully align with specific requirements, dependencies, or architectural goals, which can occasionally result in off-target solutions [69][70].

(ii)    Occasionally, ChatGPT might suggest incorrect or suboptimal solutions, particularly for complex or novel scenarios. Developers should always review and test the generated code carefully to avoid potential issues.

(iii)   While ChatGPT offers significant benefits, its integration into sensitive environments requires careful consideration. To protect proprietary data and intellectual property, organizations must implement robust security measures. Moreover, to mitigate the risk of introducing vulnerabilities, it's crucial to review and test all AI-generated code thoroughly.

These case studies reflect how companies across sectors are integrating LLMs into various stages of software engineering, showing the early promise and current limitations of AI in software development. Table 3 indicates that they can generate code and suggest improvements quickly, but they often lack the precision, ethical insight, and contextual understanding that human developers provide. Human oversight is essential to guide AI outputs, ensuring they align with real-world needs and complex project requirements. By leveraging LLMs as powerful aids, developers can focus more on creative and strategic challenges, amplifying their expertise and fostering innovation.

Table 3. Comparative Analysis

| LLMs | Key Impacts | Challenges |
|---|---|---|
| GitHub Copilot | - Productivity increase<br>- Onboarding new developers<br>- Reduction in code review load | - Code quality and security<br>- Intellectual property |
| CodeGen | - Reduced customization time<br>- Enhanced client support<br>- Agility in prototyping | - Reliability<br>- Technical debt |
| TestGen-LLM | - Increase test coverage<br>- Identify and fix bugs<br>- Accelerate development cycles | - Limited understanding of complex logic<br>- Dependency on code quality |
| ChatGPT (for Software development) | - Accelerated Coding and Development<br>- Enhanced Problem Solving and Debugging<br>- Documentation Generation<br>- Learning and Skill Development | - Context Limitations and Scope<br>- Accuracy and reliability<br>- Security and Data Privacy Concerns |

# 6. Future Directions and Research Opportunities

As Large Language Models (LLMs) continue to evolve and become more deeply integrated into software engineering (SE) processes, the future of this technology holds immense potential. However, there are several areas that still require further exploration, development, and research [71-75]. Understanding the trajectory of LLMs in SE will not only help identify their limitations but also uncover new applications and possibilities for transforming software development practices. In this section, we will explore the key future directions and research opportunities for LLMs in software engineering, ranging from technical advancements to ethical considerations and new ways to collaborate with AI models.

## 6.1. Specialization and Domain-Specific LLMs

A major area of research in the future will focus on creating more specialized LLMs tailored for specific domains within software engineering. While general-purpose LLMs like GPT-4 and Codex are highly effective across a wide range of coding tasks, they are often not optimized for niche areas such as embedded systems, real-time applications, or domain-specific languages like hardware description languages (HDL). Researchers are likely to focus on training LLMs on highly curated, domain-specific datasets, allowing these models to gain deeper expertise in specialized fields. For example, an LLM trained exclusively on medical software code or financial systems might be better equipped to understand the particular regulatory requirements, security needs, and performance constraints of these industries. Such domain-specific models could also include compliance checks that align with industry-specific standards, helping ensure that software adheres to legal and regulatory frameworks. Similarly, LLMs could be fine-tuned for particular programming languages or frameworks, providing deeper insights and optimizations tailored to those specific environments.

## 6.2. Improved Interpretability and Explainability

One of the most pressing challenges with the current generation of LLMs is their "black-box" nature, meaning they often provide answers or code suggestions without clear explanations of how or why those suggestions were made. This lack of transparency is problematic, particularly in safety-critical applications like healthcare, finance, or aerospace, where understanding the reasoning behind code is essential for ensuring security and correctness. Research in this area will likely focus on improving the interpretability and explainability of LLMs. Efforts will be made to create models that can not only generate code but also explain the rationale behind their decisions, offering developers more confidence in the accuracy and safety of the suggestions. This could involve developing new methods for LLMs to highlight the key parts of the training data or coding patterns that influenced their output. Explainable AI (XAI) frameworks that allow for deeper interrogation of LLM outputs could become more commonplace in SE environments, helping engineers better understand the suggestions provided by the models.

## 6.3. Collaborative Human-AI Programming Environments

The future of LLMs in software engineering will likely emphasize collaborative programming environments where humans and AI work together seamlessly. This will involve creating tools and platforms that promote symbiotic relationships between developers and LLMs, allowing both parties to complement each other's strengths. For instance, while LLMs excel at generating code quickly and efficiently, human developers bring contextual understanding, creativity, and ethical judgment that AI currently lacks. Research opportunities in this area include developing more intuitive, conversational interfaces for LLMs, where developers can interact with models in a fluid and iterative manner. This could involve advancements in multimodal AI, where LLMs can take into account visual inputs, such as system diagrams or wireframes, to better understand the developer's intent and provide more relevant suggestions. Similarly, AI models could be trained to adapt their suggestions based on real-time feedback from developers, improving their effectiveness over time and enabling a more interactive coding process. These collaborative environments could also include AI models acting as "pair programmers," offering continuous feedback, alternative coding approaches, and potential optimizations during the development process.

*6.4. Enhanced Debugging and Automated Bug Fixing*

One of the most promising future directions for LLMs in software engineering is their potential to revolutionize debugging and automated bug fixing. Current LLMs can already identify and suggest solutions for common errors, but future advancements may lead to more sophisticated debugging tools that can understand complex bugs in large, multi-component systems. Future research may focus on training LLMs to detect not just surface-level issues (e.g., syntax errors), but deep-rooted logical bugs, performance bottlenecks, and security vulnerabilities in more extensive codebases. For instance, LLMs of the future could autonomously analyze code dependencies and execution paths to identify the root cause of subtle issues, such as memory leaks or race conditions, which are difficult to detect manually. Moreover, they could propose multiple solutions, weigh the pros and cons of each, and recommend the best course of action, tailored to specific system constraints. Further research could explore the potential for AI to continuously monitor running systems and automatically suggest patches or improvements in real-time, reducing the need for human intervention in maintenance tasks.

*6.5. Ethical and Security Concerns*

As LLMs become more prevalent in SE, the ethical and security implications of their use will require ongoing research. For instance, as LLMs generate more and more code, questions about the ownership and licensing of that code will arise, particularly when the models are trained on publicly available, open-source projects. Who owns the code generated by AI models, and how do we ensure that it complies with existing intellectual property laws? Addressing these issues will require interdisciplinary research that involves not just software engineering but also legal scholars, ethicists, and policy makers. Another major area of concern is the security of AI-generated code. Although LLMs can detect certain types of vulnerabilities, they can also inadvertently introduce new ones. Research will need to focus on creating mechanisms that prevent LLMs from generating insecure code, particularly in mission-critical systems. There is also the risk of bias and ethical dilemmas in the datasets used to train LLMs. Models trained on biased or incomplete data may perpetuate harmful stereotypes or make inaccurate decisions, which could have significant consequences in sectors such as healthcare or criminal justice software systems. Future research will need to address ways to mitigate these risks, ensuring fairness and accountability in AI-generated code.

*6.6. Continual Learning and Model Adaptation*

As software development environments evolve, so too must the LLMs that support them. One area of research is continual learning, where LLMs can update their knowledge in real-time as they are exposed to new coding patterns, languages, or technologies. This would eliminate the need for retraining models from scratch and allow LLMs to stay relevant in dynamic environments. Future LLMs could potentially learn from real-world codebases as they evolve, adapting to new trends in development practices and adjusting their suggestions accordingly. Moreover, research into adaptive LLMs may explore models that can fine-tune themselves based on specific user needs or project contexts. For instance, a developer working on a web application might receive different types of suggestions from an LLM compared to someone working on an embedded system. Models could be fine-tuned not just for specific industries but also for individual developers, offering personalized feedback based on past interactions, coding styles, and preferred development frameworks.

*6.7. Cross-Language and Multimodal Development*

With the rise of LLMs in software engineering, there is growing interest in models that can understand and generate code across multiple programming languages. This capability would be especially useful for projects that involve integrating systems built in different languages or for teams with diverse language preferences. Research opportunities in this area include developing LLMs that are fluent in cross-language development, offering seamless transitions between languages and ensuring that code components written in different languages can work together efficiently. Additionally, multimodal LLMs that can integrate text, code, and even visual information (such as UI wireframes or architectural diagrams) offer exciting possibilities for the future of SE. These models could enable more comprehensive understanding of complex software systems, allowing developers to describe features in natural language while the LLM generates code, suggests optimizations, and aligns it with the visual or architectural elements of the project.

*6.8. Education and Training in the AI Era*

Lastly, the rise of LLMs in software engineering will have a profound impact on how future developers are trained and educated. As LLMs take over more of the rote coding tasks, the focus of SE education will likely shift toward higher-level problem-solving, system design, and ethical decision-making. Researchers will explore new pedagogical models that emphasize the collaboration between humans and AI, teaching developers not only how to code but also how to work effectively with AI tools. Future research in education will likely investigate how to integrate LLMs into software engineering curricula, ensuring that developers are well-prepared to work with AI-enhanced development tools. There will also be a need to develop new metrics for assessing coding skills, as the traditional focus on syntax and manual coding proficiency may become less relevant in a world where LLMs handle much of the low-level programming work.

## 7. Conclusion

The integration of Large Language Models (LLMs) into software engineering represents a significant turning point in how software is developed, maintained, and optimized. This article has explored the potential of LLMs to both enhance and challenge the current practices within the field of software engineering. Throughout the discussion, several important findings have emerged regarding the use of LLMs. They have proven to be game-changers across various phases of the software development lifecycle, including requirement analysis, code generation, testing, and debugging. By automating routine tasks and improving code quality, LLMs allow developers to focus on more complex and creative aspects of their work. Furthermore, LLMs can ensure consistency across large codebases and assist in maintaining legacy systems, thereby addressing technical debt effectively. However, while the potential of LLMs is vast, ethical concerns surrounding data bias, intellectual property, and job displacement must be carefully managed. The computational costs associated with training and deploying these large-scale models can also be prohibitive, particularly for smaller organizations. In light of these findings, it is clear that LLMs are not merely a product of overhyped marketing; they represent a profound shift in how software is engineered. They should be seen as powerful tools that augment human capabilities rather than replace them. Human oversight remains crucial for ensuring that AI-generated code aligns with project goals, is secure, and is free from biases. Therefore, the verdict is that LLMs indeed are game-changers in software engineering, but their true potential can only be unlocked when combined with human expertise and ethical safeguards. For developers and organizations, embracing the rise of LLMs is not just a choice but a strategic imperative. Developers must become familiar with how LLMs can assist in coding, testing, debugging, and maintenance while continuing to refine their higher-level skills such as system design and ethical decision-making. Organizations should invest in integrating LLMs into their development environments, starting with pilot projects to gauge effectiveness, as this can reduce development costs, accelerate time-to-market, and enhance software quality. Educational institutions, too, should revise their software engineering curricula to prepare the next generation of developers for the future of AI-driven development.

The impact of this article extends beyond simply presenting the advantages and challenges of LLMs in software engineering; it provides a balanced and nuanced perspective that allows stakeholders to make informed decisions about adopting these technologies. By highlighting real-world case studies, technical strengths, ethical considerations, and future research opportunities, the article contributes to the growing discourse on AI-driven development tools and their place in the future of software engineering. Ultimately, it serves as a guide for developers, organizations, and researchers, helping them understand how LLMs can enhance workflows and the skills needed to remain competitive in an AI-driven landscape. As LLMs continue to evolve, their integration into software engineering practices will redefine what is possible in software development, pushing the boundaries of automation, creativity, and collaboration. Thus, this article offers a foundational understanding of how LLMs are poised to change the software engineering landscape, encouraging stakeholders to embrace these tools thoughtfully and strategically. In conclusion, LLMs hold the potential to significantly disrupt and enhance the software engineering process, and as developers and organizations adapt to these changes, they will find themselves at the forefront of a new era in software development—one that is faster, more efficient, and more collaborative than ever before.

# REFERENCES

[1] Hironori Washizaki, eds. Guide to the Software Engineering Body of Knowledge (SWEBOK Guide), Version 4.0. IEEE Computer Society, 2024. www.swebok.org.

[2] Roger S Pressman. Software Engineering: A Practitioner's Approach. 5th Edition, McGraw-Hill Higher Education, 2001, ISBN- 0073655783.

[3] Pankaj Jalote. Software Engineering: A Precise Approach. Wiley, 2010, ISBN-9788126523115.

[4] MA Haque, Nesar Ahmad. 2022. Key Issues in Software Reliability Growth Models. Recent Advances in Computer Science and Communications, vol. 15(5), pp. 741-747.

[5] Xinyi Hou, Yanjie Zhao, Yue Liu, Zhou Yang, Kailong Wang, Li Li, Xiapu Luo, David Lo, John Grundy, and Haoyu Wang. 2024. Large Language Models for Software Engineering: A Systematic Literature Review. ACM Transactions on Software Engineering and Methodology. https://doi.org/10.1145/3695988.

[6] Ipek Ozkaya. 2023. Application of large language models to software engineering tasks: Opportunities, risks, and implications. IEEE Software, vol. 40, no. 3, pp. 4-8.

[7] Sanka Rasnayaka, Guanlin Wang, Ridwan Shariffdeen, and Ganesh Neelakanta Iyer. 2024. An Empirical Study on Usage and Perceptions of LLMs in a Software Engineering Project. Proceedings of the 1st International Workshop on Large Language Models for Code, Pages 111 – 118. https://doi.org/10.1145/3643795.3648379.

[8] Yao Li, Tao Zhang, Xiapu Luo, Haipeng Cai, Sen Fang, Dawei Yuan. 2023. Do Pretrained Language Models Indeed Understand Software Engineering Tasks?. IEEE Transactions on Software Engineering, vol. 49, no. 10, pp. 4639-4655.

[9] Zhijie Liu, Yutian Tang, Xiapu Luo, Yuming Zhou, Liang Feng Zhang. 2024. No Need to Lift a Finger Anymore? Assessing the Quality of Code Generation by ChatGPT. IEEE Transactions on Software Engineering, vol. 50, no. 6, pp. 1548-1584.

[10] Juyong Jiang, Fan Wang, Jiasi Shen, Sungju Kim, Sunghun Kim. 2024. A Survey on Large Language Models for Code Generation. arXiv preprint arXiv:2406.00515v1.

[11] Artur Tarassow. 2023. The potential of llms for coding with low-resource and domain-specific programming languages. arXiv preprint arXiv:2307.13018.

[12] June Sallou, Thomas Durieux, Annibale Panichella. 2024. Breaking the silence: the threats of using LLMs in software engineering. Proceedings of the 2024 ACM/IEEE 44th International Conference on Software Engineering: New Ideas and Emerging Results, Pages 102 - 106.

[13] Ashish Vaswani et al. 2017. Attention Is All You Need. Advances in Neural Information Processing Systems (NIPS 2017), vol. 30, Long Beach, CA, USA.

[14] TB Brown et al. 2020. Language Models are Few-Shot Learners. Proc. of the 34th Int. Conf. on Neural Information Processing Systems (NIPS '20), Article 159, 1877–1901.

[15] Jeremy Howard, Sebastian Ruder. 2018. Universal language model fine-tuning for text classification. 56th Annual Meeting of the Association for Computational Linguistics (Long Papers), pages 328–339, Melbourne, Australia.

[16] Long Ouyang et al. 2022. Training language models to follow instructions with human feedback. 36th Conf. on Neural Information Processing Systems, vol.35, 27730–27744.

[17] MA Haque, S Li. 2024. Exploring ChatGPT and its Impact on Society. AI and Ethics. https://doi.org/10.1007/s43681-024-00435-4.

[18] Mohammed Lubbad. 2023. GPT-4 Parameters: Unlimited guide NLP's Game-Changer. Medium (March 19, 2023). Available online: https://medium.com/@mlubbad/the-ultimate-guide-to-gpt-4-parameters-everything-you-need-to-know-about-nlps-game-changer-109b8767855a.

[19] WX Zhao et al. 2023. A Survey of Large Language Models. arXiv preprint arXiv:2303.18223v11.

[20] Ross Taylor, Marcin Kardas, Guillem Cucurull, Thomas Scialom, Anthony Hartshorn, Elvis Saravia, Andrew Poulton, Viktor Kerkez, Robert Stojnic. 2022. Galactica: A large language model for science. arXiv preprint arXiv:2211.09085.

[21] Hugo Touvron et al. 2023. Llama: Open and efficient foundation language models. arXiv preprint arXiv:2302.13971.

[22] Yupeng Chang et al. 2024. A survey on evaluation of large language models. ACM Transactions on Intelligent Systems and Technology, Vol 15, Issue 3, Article No. 39, Pages 1 - 45.

[23] Simiao Zhang, Jiaping Wang, Guoliang Dong, Jun Sun, Yueling Zhang, Geguang Pu. 2024. Experimenting a new programming practice with llms. arXiv preprint arXiv:2401.01062.

[24] Rasha Ahmad Husein, Hala Aburajouh, Cagatay Catal. 2025. Large language models for code completion: A systematic literature review. Computer Standards & Interfaces, vol. 92, 103917.

[25] M. Welsh. 2023. The end of programming. Communications of the ACM, vol. 66, no. 1, pp. 34–35. https://doi.org/10.1145/3570220.

[26] Nalin Wadhwa, Jui Pradhan, Atharv Sonwane, Surya Prakash Sahu, Nagarajan Natarajan, Aditya Kanade, Suresh Parthasarathy, Sriram Rajamani. 2023. Frustrated with code quality issues? llms can help!. arXiv preprint arXiv:2309.12938.

[27] Zejun Wang, Jia Li, Ge Li, Zhi Jin. 2023. Chatcoder: Chat-based refine requirement improves llms' code generation. arXiv preprint arXiv:2311.00272.

[28] Ryan Yen, Jiawen Zhu, Sangho Suh, Haijun Xia, Jian Zhao. 2023. Coladder: Supporting programmers with hierarchical code generation in multi-level abstraction. arXiv preprint arXiv:2310.08699.

[29] K Huang et al. 2023. An empirical study on fine-tuning large language models of code for automated program repair. Proceedings 38th IEEE/ACM Int. Conf. Automated Softw. Eng. (ASE), pp. 1162-1174.

[30] MA Haque, S Li. 2023. The Potential Use of ChatGPT for Debugging and Bug Fixing. EAI Endorsed Transactions on AI and Robotics, vol. 2(1), e4.

[31] Fernando Vallecillos Ruiz, Anastasiia Grishina, Max Hort, Leon Moonen. 2024. A novel approach for automatic program repair using round-trip translation with large language models. arXiv preprint arXiv:2401.07994.

[32] S Kang, J Yoon, N Askarbekkyzy, S Yoo. 2024. Evaluating Diverse Large Language Models for Automatic and General Bug Reproduction. IEEE Transactions on Software Engineering, vol. 50, no. 10, pp. 2677-2694.

[33] Zhiyu Fan, Xiang Gao, Martin Mirchev, Abhik Roychoudhury, Shin Hwei Tan. 2023. Automated Repair of Programs from Large Language Models. 2023 IEEE/ACM 45th International Conference on Software Engineering (ICSE), Melbourne, Australia, pp. 1469-1481.

[34] AZH Yang, Ruben Martins, Claire Le Goues, VJ Hellendoorn. 2024. Large Language Models for Test-Free Fault Localization. Proceedings of the IEEE/ACM 46th Int. Conference on Software Engineering (ICSE '24), Article 17, 1–12.

[35] Junjie Wang, Yuchao Huang, Chunyang Chen, Zhe Liu, Song Wang, Qing Wang. 2024. Software Testing With Large Language Models: Survey, Landscape, and Vision. IEEE Transactions on Software Engineering, vol. 50, no. 4, pp. 911-936.

[36] Max Schafer, Sarah Nadi, Aryaz Eghbali, Frank Tip. 2024. An Empirical Evaluation of Using Large Language Models for Automated Unit Test Generation. IEEE Transactions on Software Engineering, vol. 50, no. 1, pp. 85-105.

[37] Tengfei Xue, Xuefeng Li, Tahir Azim, Roman Smirnov, Jianhui Yu, Arash Sadrieh, Babak Pahlavan. 2024. Multi-Programming Language Ensemble for Code Generation in Large Language Model. arXiv preprint arXiv:2409.04114.

[38] Ben Athiwaratkun e. al. 2023. Multi-lingual evaluation of code generation models. arXiv preprint arXiv:2210.14868.

[39] Jiyang Zhang, Pengyu Nie, Junyi Jessy Li, Milos Gligoric. 2023. Multilingual Code Co-evolution using Large Language Models. Proceedings of the 31st ACM Joint European Software Engineering Conference and Symposium on the Foundations of Software Engineering (ESEC/FSE 2023), 695–707.

[40] Qiwei Peng, Yekun Chai, Xuhong Li. 2024. Humaneval-xl: A multilingual code generation benchmark for cross-lingual natural language generalization. Proceedings of the 2024 Joint International Conference on Computational Linguistics, Language Resources and Evaluation, pages 8383–8394, Torino, Italia.

[41] Atsushi Shirafuji, Yusuke Oda, Jun Suzuki, Makoto Morishita, Yutaka Watanobe. 2023. Refactoring Programs Using Large Language Models with Few-Shot Examples. 30th Asia-Pacific Software Engineering Conference (APSEC-23), Seoul, South Korea, pp. 151-160.

[42] Shu Ishida et al. 2024. LangProp: A code optimization framework using Large Language Models applied to driving. arXiv preprint arXiv:2401.10314.

[43] AD Porta et al. 2024. Using Large Language Models to Support Software Engineering Documentation in Waterfall Life Cycles: Are We There Yet?. Proceedings of the 4th National Conference on Artificial Intelligence, organized by CINI, May 29-30, Naples, Italy.

[44] Lenz Belzner, Thomas Gabor, and Martin Wirsing. 2023. Large Language Model Assisted Software Engineering: Prospects, Challenges, and a Case Study. In Bridging the Gap Between AI and Reality: First International Conference, AISoLA 2023, Crete, Greece, 355–374.

[45] Haolin Jin, Linghan Huang, Haipeng Cai, Jun Yan, Bo Li, Huaming Chen. 2024. From LLMs to LLM-based Agents for Software Engineering: A Survey of Current, Challenges and Future. arXiv preprint arXiv:2408.02479.

[46] Federico Errica, Giuseppe Siracusano, Davide Sanvito, Roberto Bifulco. 2024. What Did I Do Wrong? Quantifying LLMs' Sensitivity and Consistency to Prompt Engineering. arXiv preprint arXiv:2406.12334v1.

[47] Yuchen Xia, Jiho Kim, Yuhan Chen, Haojie Ye, Souvik Kundu, Cong Hao, Nishil Talati. Understanding the Performance and Estimating the Cost of LLM Fine-Tuning. arXiv preprint arXiv:2408.04693.

[48] MC Rillig, Marlene Agerstrand, Mohan Bi, KA Gould, Uli Sauerland. 2023. Risks and benefits of large language models for the environment", Environmental Science & Technology 57, 9, 3464–3466.

[49] C Tantithamthavorn, J Cito, H Hemati, S Chandra. 2023. Explainable AI for SE: Experts' interviews, challenges, and future directions. IEEE Software, vol. 40, no. 4.

[50] Bonan Kou, Shengmai Chen, Zhijie Wang, Lei Ma, and Tianyi Zhang. 2023. Do Large Language Models Pay Similar Attention Like Human Programmers When Generating Code?. arXiv preprint arXiv:2306.01220.

[51] Neil Perry, Megha Srivastava, Deepak Kumar, Dan Boneh. 2023. Do Users Write More Insecure Code with AI Assistants?". ACM SIGSAC Conference on Computer and Communications Security (CCS '23), 2785–2799.

[52] Christin Kirchhubel and Georgina Brown. 2024. Intellectual property rights at the training, development and generation stages of Large Language Models. In Proceedings of the Workshop on Legal and Ethical Issues in Human Language Technologies @ LREC-COLING 2024, pages 13–18, Torino, Italia. ELRA and ICCL.

[53] Junfeng Jiao, Saleh Afroogh, Yiming Xu, Connor Phillips. 2024. Navigating LLM Ethics: Advancements, Challenges, and Future Directions. arXiv preprint arXiv:2406.18841

[54] Nestor Maslej, Loredana Fattorini, Raymond Perrault, Vanessa Parli, Anka Reuel, Erik Brynjolfsson, John Etchemendy, Katrina Ligett, Terah Lyons, James Manyika, Juan Carlos Niebles, Yoav Shoham, Russell Wald, Jack Clark. 2024. The AI Index 2024 Annual Report. AI Index Steering Committee, Institute for Human-Centered AI, Stanford University, CA.

[55] Kyle Daigle & GitHub Staff. 2024. Survey: The AI wave continues to grow on software development teams. GitHub Blog. Available online: https://github.blog/news-insights/research/survey-ai-wave-grows/#key-survey-findings.

[56] S Imai. 2022. Is GitHub copilot a substitute for human pair-programming? an empirical study. Proceedings of the ACM/IEEE 44th International Conference on Software Engineering (ICSE '22), Pittsburgh, Pennsylvania, 319–321.

[57] M Jaworski, D Piotrkowski. 2023. Study of software developers' experience using the Github Copilot Tool in the software development process. arXiv preprint arXiv:2301.04991.

[58] Sida Peng, Eirini Kalliamvakou, Peter Cihon, Mert Demirer. 2023. The Impact of AI on Developer Productivity: Evidence from GitHub Copilot. arXiv preprint arXiv:2302.06590.

[59] Yujia Fu, Peng Liang, Amjed Tahir, Zengyang Li, Mojtaba Shahin, Jiaxin Yu, Jinfu Chen. 2023. Security Weaknesses of Copilot Generated Code in GitHub. arXiv preprint arXiv:2310.02059.

[60] Erik Nijkamp, Bo Pang, Hiroaki Hayashi, Lifu Tu, Huan Wang, Yingbo Zhou, Silvio Savarese, Caiming Xiong. 2023. CodeGen: An Open Large Language Model for Code with Multi-Turn Program Synthesis. arXiv preprint arXiv:2203.13474v5.

[61] Aman Madaan et al. 2023. Learning Performance-Improving Code Edits. Available online: https://dl4c.github.io/assets/pdf/papers/28.pdf.

[62] Nadia Alshahwan, Jubin Chheda, Anastasia Finegenova, Beliz Gokkaya, Mark Harman, Inna Harper, Alexandru Marginean, Shubho Sengupta, Eddy Wang. 2024. Automated Unit Test Improvement using Large Language Models at Meta. arXiv preprint arXiv:2402.09171.

[63] Oluwadamisi Samual. 2024. How to Use AI to Automate Unit Testing with TestGen-LLM and Cover-Agent. FreeCodeCamp (June 3, 2024). Available online: https://www.freecodecamp.org/news/automated-unit-testing-with-testgen-llm-and-cover-agent/.

[64] Arifa Islam Champa, Md Fazle Rabbi, Costain Nachuma, and Minhaz F. Zibran. 2024. ChatGPT in Action: Analyzing Its Use in Software Development. 2024 IEEE/ACM 21st International Conference on Mining Software Repositories (MSR), Lisbon, Portugal, pp. 182-186.

[65] Dae-Kyoo Kim, Jingshu Chen, Hua Ming, and Lunjin Lu. 2023. Assessment of ChatGPT's Proficiency in Software Development. Proceedings of CSCE-2023, Las Vegas, USA, pp. 2637-2644.

[66] Zhiqiang Yuan, Yiling Lou, Mingwei Liu, Shiji Ding, Kaixin Wang, Yixuan Chen, Xin Peng. 2023. No More Manual Tests? Evaluating and Improving ChatGPT for Unit Test Generation. arXiv preprint arXiv:2305.04207.

[67] N Marques, RR Silva, J Bernardino. 2024. Using ChatGPT in Software Requirements Engineering: A Comprehensive Review. Future Internet, 16(6):180.

[68] Ranim Khojah, Mazen Mohamad, Philipp Leitner, Francisco Gomes de Oliveira Neto. 2024. Beyond Code Generation: An Observational Study of ChatGPT Usage in Software Engineering Practice. arXiv preprint arXiv:2404.14901v2.

[69] MA Haque. 2023. A Brief Analysis of 'ChatGPT' – A Revolutionary Tool Designed by OpenAI. EAI Endorsed Transactions on AI and Robotics, vol. 1(1), e15.

[70] W Rahmaniar. 2024. ChatGPT for Software Development: Opportunities and Challenges. IT Professional, vol. 26, no. 03, pp. 80-86.

[71] Daniel Russo. 2024. Navigating the Complexity of Generative AI Adoption in Software Engineering. ACM Trans. Softw. Eng. Methodol. 33, 5, Article 135 (June 2024), 50 pages. https://doi.org/10.1145/3652154.

[72] Antony Drake. 2024. The Future of Software Engineering: LLMs and Beyond. Comet (February 28, 2024). Available online: https://www.comet.com/site/blog/the-future-of-software-engineering-llms-and-beyond/.

[73] Joseph Monti. 2024. The Future of AI-Driven Software Development. Medium (Mar 8, 2024). Available online: https://joemonti.org/the-future-of-ai-driven-software-development-0dec24759a71.

[74] Jaakko Sauvola, Sasu Tarkoma, Mika Klemettinen, Jukka Riekki, David Doermann. 2024. Future of software development with generative AI. Automated Software Engineering 31, 26. https://doi.org/10.1007/s10515-024-00426-z.

[75] Valerio Terragni, Partha Roop, Kelly Blincoe. 2024. The Future of Software Engineering in an AI-Driven World. arXiv preprint arXiv:2406.07737.